\documentclass[aps,prl,a4paper,reprint,amsmath,amssymb]{revtex4-2}
\usepackage{graphicx}
\usepackage{color}
\usepackage{hyperref} 
\begin{document}

\title{Edge State Selective Measurement of Quantum Hall Dispersions}
\author{Henok Weldeyesus}
\email{henok.weldeyesus@unibas.ch}
\affiliation{Departement Physik, University of Basel, Klingelbergstrasse 82, CH-4056 Basel, Switzerland}
\author{T.~Patlatiuk}
\affiliation{Departement Physik, University of Basel, Klingelbergstrasse 82, CH-4056 Basel, Switzerland}
\author{Q.~Chen}
\altaffiliation[Present address: ]{Kavli Institute for Theoretical Sciences, University of Chinese Academy of Sciences, Beijing 100190, China}
\affiliation{Departement Physik, University of Basel, Klingelbergstrasse 82, CH-4056 Basel, Switzerland}
\author{C.~P.~Scheller}
\affiliation{Departement Physik, University of Basel, Klingelbergstrasse 82, CH-4056 Basel, Switzerland}
\author{A.~Yacoby}
\affiliation{Department of Physics, Harvard University, Cambridge, Massachusetts 02138, USA}
\author{L.~N.~Pfeiffer}
\affiliation{Department of Electrical Engineering, Princeton University, Princeton, New Jersey 08544, USA}
\author{K.~W.~West}
\affiliation{Department of Electrical Engineering, Princeton University, Princeton, New Jersey 08544, USA}
\author{D.~M.~Zumb\"uhl}
\email{dominik.zumbuhl@unibas.ch}
\affiliation{Departement Physik, University of Basel, Klingelbergstrasse 82, CH-4056 Basel, Switzerland}

\begin{abstract}
Edge states reflect the key physical properties yet are difficult to probe individually, particularly when several states are present at an edge. We present momentum resolved tunneling spectroscopy between a quantum well and a quantum wire to extract the dispersions of the quantum Hall edge states. Momentum and energy selective tunneling allows to separately address the different states even if they are spatially overlapping. This delivers the edge state velocities over broad ranges of magnetic field and density, in excellent agreement with a hard-wall model. This technique provides a basis for future edge state selective spectroscopy on quantum materials.  
\end{abstract}
\maketitle

\begin{figure}[htb!]
    \centering
    \includegraphics[width=\columnwidth]{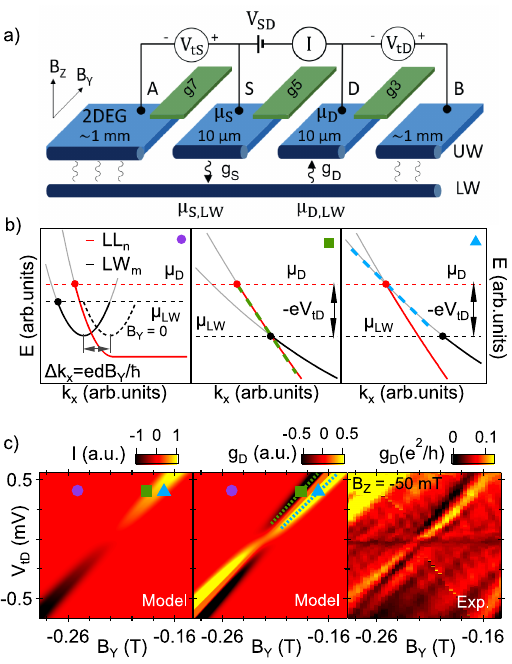}
    \caption{ \textbf{Bias voltage tunneling spectroscopy} (a) Sample schematic for the velocity measurement. (b) Schematic of the dispersion for tunneling from a Landau level (LL) to a lower wire mode, at fixed tunnel junction voltage $V_{tD}$ and varying magnetic fields from left to right when going through the panels. The left panel corresponds to a non-resonant condition. The second and third panels show a magnified view at two values of $B_Y$, corresponding to tracing out $LL_n$ with the Fermi-point of the $LW_m$ and vice versa. (c) Simulated current and conductance (left and center panel) and corresponding conductance measurement, used to extract the velocities for a given LL and out-of-plane magnetic field $B_Z$. Each such measurement typically provides two mode velocities, a fast and a slow one. The colored markers in the first two panels mark the positions of the situations depicted in (b). }
     \label{fig:1}
\end{figure}

 Edge states appearing on the boundary of topological insulators \cite{Hasan2010} and topological superconductors \cite{Qi2011} are directly tied to the underlying bulk properties through bulk-boundary correspondence. 
To probe key properties such as the dispersions or the associated velocities of such edge states is, however, a more difficult task. Spectroscopic measurements such as angle resolved photoemission can play a crucial role in the investigation of the edge state properties, and are often used to study surface and edge states of topological systems where the location of these states is exposed and accessible to optical measurements \cite{Lisi2020}. Consequently they are not applicable to quantum wells or buried structures in which many studies are performed. 

Previous characterizations of quantum Hall edge state velocities in GaAs -- one of the most studied material systems for the integer and fractional quantum Hall effects due to its ultra-high mobility -- relied on time-of-flight measurements of edge magnetoplasmons \cite{Kamata2010, Kumada2011, Kataoka2016, Lin2021}. This technique was also applied to topological phases like the quantum anomalous Hall effect \cite{Martinez} and the quantum spin Hall effect \cite{Gourmelon}, but requires challenging ultrafast time-resolved setups. We note that the magnetoplasmon velocity is only vaguely related to the edge state velocity since it is a collective mode rather than a single edge state. Alternatively, quantum interferometry \cite{McClure2009, Gurman2016, Ji2003, Neder2006,  Nakamura2020, Roeoesli2021} was also employed. This requires sufficiently small structures to keep the edge states coherent and is generally masked by competing Coulomb charging effects  \cite{Halperin2011, Nakamura2019}.  

An important aspect of the investigation of edge states is the ability to selectively study their properties. Momentum-resolved tunneling spectroscopy presents a well established method to  characterize the properties of electronic states in (partially) translation invariant systems. Conservation of momentum automatically ensures individual addressability of the edge states if these have different momenta. Cleaved edge overgrowth devices \cite{Pfeiffer1993} in various geometries \cite{Schedelbeck1997,Grayson1996} and gate-defined quantum wires \cite{Kardynal1997} are most prominent. They were used to investigate 1D dispersions \cite{Auslaender2002} and Luttinger liquid physics such as spin-charge separation \cite{Auslaender2005, Jompol2009} and charge fractionalization \cite{Steinberg2008}. 
Quantum Hall edge states in such systems were also investigated with similar spectroscopy techniques, for example edge state hybridization \cite{Kang2000,Patlatiuk2018}, edge confinement \cite{Huber2002,Huber2005}, edge state position evolution \cite{Patlatiuk2018} and finally also their wave functions \cite{Patlatiuk2020}.

In this Letter, we obtain the dispersions and velocities of edge states formed in a 2D electron gas (2DEG) in a cleaved edge overgrowth double quantum well structure. We probe the dispersion of a selected edge state by choosing its momentum with an in-plane magnetic field $B_Y$. We pursue two approaches: at constant density, we probe the edge state evolution with bias voltage from small to large perpendicular magnetic field $B_Z$ which controls the Landau levels (LLs). Alternatively, we fix $B_Z$ and vary the density below a gate, following the lowest five LLs  until their depletion. We compare the extracted velocities to a single-particle model using hard-wall edge confinement and find good agreement with the measurements. 

A diagram of the sample and measurement scheme is shown in Fig.\,\ref{fig:1} (a). A 2DEG with an adjacent upper wire (UW) is tunnel-coupled to a lower wire (LW) directly below it. Surface gates g7, g5, and g3 are used to deplete the 2DEG and UW, forming separate $10\,\mathrm{\mu m}$ sized regions, as shown. Current flows from a source ohmic contact (S) with the chemical potential $\mu_S$ through the 2DEG and the UW, tunneling to the lower wire (thick wiggled arrow), propagating under g5, and finally tunneling back to the upper system into the drain (D), which is grounded.  We apply a source-drain voltage $V_{SD}$ and a small AC voltage of 6\,$\mathrm{\mu V}$ at low frequencies ($\sim$ 17\,Hz) to obtain the differential conductance $g$ via standard lock-in measurements. In addition, we measure the voltage drop on the tunnel junctions, the source tunneling voltage $V_{tS}$ and the drain tunneling voltage $V_{tD}$, with two voltage probes $A$ and $B$. This is needed because the voltage drop has a highly nonlinear dependence on the applied bias voltage $V_{SD}$, and cannot be accounted for without a separate measurement. Here we will focus on the data from the drain junction. More details on the voltage measurement and complementary data from the source junction can be found in the supplementary \cite{supplementary}. 

We apply an in-plane magnetic field $B_Y$ and an out-of-plane magnetic field $B_Z$, as indicated in Fig.\,\ref{fig:1}(a). The spectroscopy field $B_Y$ is parallel to the 2DEG and perpendicular to the plane spanned by the quantum wires, and therefore provides a momentum kick $\Delta k_x=eB_Yd/\hbar$ to tunneling electrons \cite{Auslaender2002}. Here, $e$ is the electron charge, $d$ is the center-to-center distance between the quantum wells, and $\hbar$ is the reduced Planck constant. This allows us to perform spectroscopy as a function of momentum by tuning $B_Y$, 
and therefore provides the ability to selectively tunnel to only a single edge state even when many are present at that edge. Because of this momentum conserving tunneling process, this spectroscopy can also be performed at filling fractions corresponding to a conductive 2D bulk, where other techniques are not applicable.  
 
The field $B_Z$, applied perpendicular to the 2DEG, induces the LLs. Throughout this Letter we only refer to the spin-degenerate LL since large magnetic fields are needed to resolve the spins \cite{Patlatiuk2018}. In addition to the Landau quantization, $B_Z$ provides additional momentum  when tunneling, related to the displacement in the y-direction of the upper and lower system, such that the total momentum shift becomes $\Delta k_x= eB_Yd/\hbar + eB_Z\Delta y/\hbar$ \cite{Patlatiuk2018, Patlatiuk2020}. This provides independent control of the momentum kick and the LLs. By applying a bias voltage, the dispersions in the upper and lower systems are shifted in the energy direction.  The densities are assumed to be independent of the applied bias voltage \cite{Boese2001}.  The measured tunnel current is enhanced when $\Delta k_x$ and $V_{SD}$ are adjusted such that filled states of one dispersion align with empty states in the other dispersion \cite{Auslaender2000}.

We now describe the extraction of the dispersion relations from the tunneling spectroscopy. We start by showing the calculated dispersion relation of a LL $\mathrm{LL_n}$ (red) and a lower wire mode $\mathrm{LW_m}$ (black) for three different values of the spectroscopy field at fixed voltage $V_{tD}$ across the drain junction, see Fig.\,\ref{fig:1}(b). The states above the chemical potential (horizontal dashed lines) are not populated with electrons and are shown in gray. The first panel shows the off-resonant case, where no matching is present (only filled states overlap each other close to the bottom of the LW dispersion). The second panel shows the matching of the LW Fermi point with the filled states in the LL dispersion at a specific value of $B_Y$. This causes the current flow to turn on and remain on while empty and filled states are overlapping at the intersection of the two dispersions. When the LL Fermi point starts to shift off the LW dispersion (right panel), the current shuts down. 

The current turns on or off precisely when a Fermi point overlaps with the other dispersion, see the first panel in Fig.\,\ref{fig:1} (c), which results in a corresponding peak or dip in the differential conductance, see Fig.\,\ref{fig:1}(c). The peak and the dip indicate where one dispersion is mapped using the Fermi point of the other one, thus allowing us to separately extract each dispersion, i.e. the dispersions of the LL and the LW modes. In the experiment, it is not always easy to identify such pairs of differential conductance peaks since other features such as finite size effects \cite{Tserkovnyak2002} or other modes can interfere. A relatively clean example is shown in Fig.\,\ref{fig:1}(c)(third panel). The dispersions appear as lines here since the curvature is negligible given the applied bias is relatively small compared to the Fermi energy. The slope of these lines give the velocity of the corresponding modes,
\begin{equation} \label{eq2}
v = \frac{1}{\hbar} \frac{d E}{d k_x} = \frac{1}{d} \frac{d V_{tD}}{d B_Y},
\end{equation}
where $E$ and $k_x$ are the energy and momenta of the states.

\begin{figure}[htb!]
	\centering
	\includegraphics[width=\columnwidth]{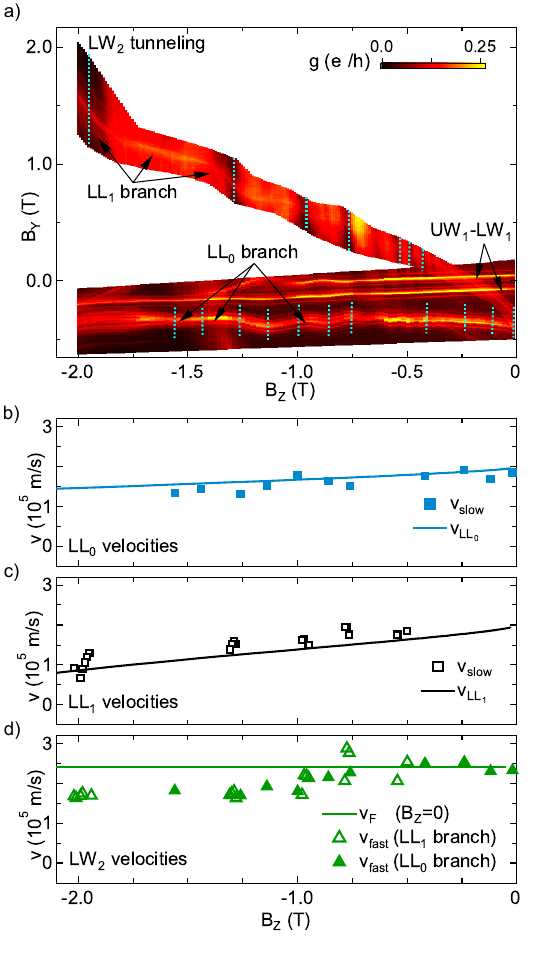}
	\caption{\textbf{Magnetic field evolution of velocities} (a) Zero bias tunnel map showing resonances as a function of out-of-plane field $B_Z$ and spectroscopy field $B_Y$. Bias spectroscopy measurements are performed at each of the blue dashed vertical lines crossing both LL branches. (b, c) Slow velocities measured for both LL$_0$ (b), and LL$_1$ branches (c) together with a hard wall model (solid lines) (d) The corresponding fast velocities together with the calculated $\mathrm{LW_2}$ Fermi velocity  (solid green line), assumed to be $B_Z$ independent. }
	\label{fig:2}
\end{figure}

To study the evolution of LL velocities with the out-of-plane magnetic field we first identify the position of the tunneling resonances in the $B_Y$\nobreakdash-$B_Z$ plane. These correspond to the matching of the LL and LW Fermi-points at constant bias voltage, and are shown in Fig.\,\ref{fig:2} (a) for $\mathrm{LL_0}$ and $\mathrm{LL_1}$ tunneling to $\mathrm{LW_2}$, see Ref.~\cite{Patlatiuk2018} for the detailed identification of the various modes. Then, $B_Z$ is fixed and the bias spectroscopy, as shown in  Fig.\,\ref{fig:1} (c), is performed in a small range of spectroscopy field $B_Y$ in the vicinity of the tunneling resonances. The values of $B_Z$ where the spectroscopy is carried out are indicated by blue vertical lines in Fig.\,\ref{fig:2} (a). These positions were chosen due to the increased contrast with respect to the background conductance.$\mathrm{LW_2}$

The bias spectroscopy delivers always two velocities which need to be properly assigned to the LL or the LW. The LW mode is quite strongly confined and therefore relatively weakly affected by a moderate $B_Z$ field, while the LLs increase their energy linearly with $B_Z$ and the edge state velocities are reduced as the bulk LL approaches the Fermi level. We obtain a fast and a slow value from both LL branches (see Fig.\,\ref{fig:2}(a)), delivering four different velocities. Then, we identify the two modes with the stronger $B_Z$ dependence as $\mathrm{LL_0}$ and $\mathrm{LL_1}$, plotted separately in Fig.\,\ref{fig:2} (b,c), and the remaining two modes as two separate measurements of the $\mathrm{LW_2}$ velocities, see Fig.\,\ref{fig:2} (d). 

We find good agreement for the LL velocities with a hard wall model (see supplementary) applicable here for the sharp cleaved edge potential, see solid curves. We note that 2DEG density is enhanced by about 50\% at the edge compared to the measured bulk density. This is due to the presence of the ionized donors on the cleaved edge side as previously reported \cite{Patlatiuk2020,Patlatiuk2018}. The $\mathrm{LW_2}$ velocities turn out to be the fastest ones extracted here, and we see good agreement between the values from both LL branches. Further, they also agree well with the calculated Fermi velocity (solid green line) from the densities in these modes as obtained from low-B spectroscopy \cite{Patlatiuk2018}. As $B_Z$ is increased above $\sim1$\,T, a slight reduction of the measured $\mathrm{LW_2}$ velocities is observed, consistent with incipient magnetic depopulation of the mode (not included in the model).  

\begin{figure}[htb!]
	\centering
	\includegraphics[width=\columnwidth]{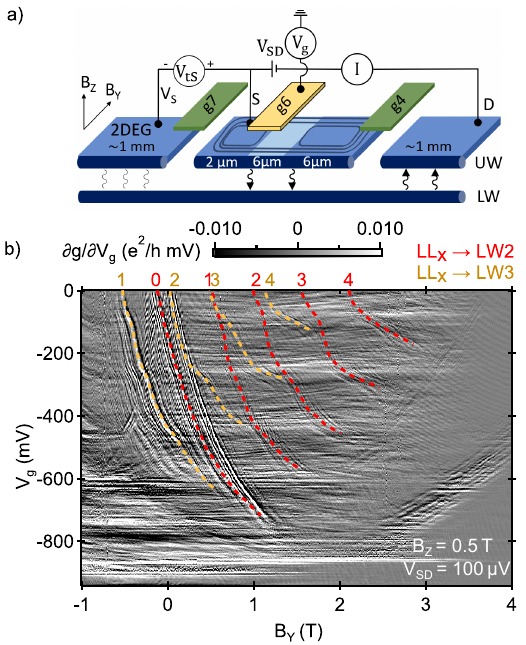}
	\caption{\textbf{Density dependent spectroscopy} (a) Setup for the density dependent measurement. The conductance is measured on the source-junction, on which a gate (g6) is used to locally change the density. (b) Zero bias tunnel map showing resonances as a function of spectroscopy field $B_Y$ and gate voltage $V_{g6}$ performed at $B_Z=0.5\,T$. The red and orange curves correspond to tunneling between the LL and the $\mathrm{LW_2}$ or $\mathrm{LW_3}$, respectively. }
	\label{fig:3}
\end{figure}

We also perform a second experiment at constant bias voltage, and instead extract the velocities at constant $B_Z$ as a function of density controlled with a gate voltage. As before, we will observe increased conductance when matching the Fermi points of the LW modes with the LL dispersion or vice versa. Instead of shifting the dispersions in energy, the reduction in the density of both the 2DEG and the wire modes causes a change in their Fermi wavenumber $k_F$, which leads to a shift of the resonance position in the ($V_{g}$, $B_Y$) plane (see supplementary). The measurement configuration is shown in Fig. \ref{fig:3} (a) and the gate voltage $V_{g}$ is applied to gate $\mathrm{g_6}$. The adjacent gates $\mathrm{g_7}$ and $\mathrm{g_4}$ are defining the source tunnel junction by depleting the 2DEG and upper wire. We fix the out-of-plane field $B_Z$ at 0.5\,T, and then record the tunneling conductance as a function of spectroscopy field $B_Y$ and top-gate voltage $V_{g}$. In principle this experiment can be performed at zero bias, but a fixed, small bias voltage of $V_{SD}\sim100\,\mu$V is applied to avoid the zero-bias anomaly \cite{Tserkovnyak2003,Scheller2014a}. This simplified measurement scheme avoids the voltage measurement of Fig.~\ref{fig:1}(a) since the position of the resonance does not strongly depend on bias voltage. 

The measurement is shown in Fig. \ref{fig:3} (b), where a derivative with respect to gate voltage was taken to remove a smooth background stemming from the ungated section of the source region. Two sets of resonances are shown in red and yellow, denoting tunneling to $\mathrm{LW_2}$ and $\mathrm{LW_3}$, respectively. The mode assignments are apparent from a larger $B_Z-B_Y$ map shown in Ref.~\cite{Patlatiuk2018} and the mode densities are extracted in the supplementary \cite{supplementary}. These resonances are only visible up to certain gate voltages, where they vanish, marking the depletion of that LL and the corresponding edge state. In addition to the marked resonances, replicas due to finite size effects of the 6\,$\mu$m gate, are seen to follow in parallel \cite{Tserkovnyak2002}. Where a LL crosses a wire mode in the upper system, hybridization gaps are observed \cite{Patlatiuk2020}, which is discussed in more detail in the supplementary \cite{supplementary}.      

\begin{figure}[htb!]
	\centering
	\includegraphics[width=\columnwidth]{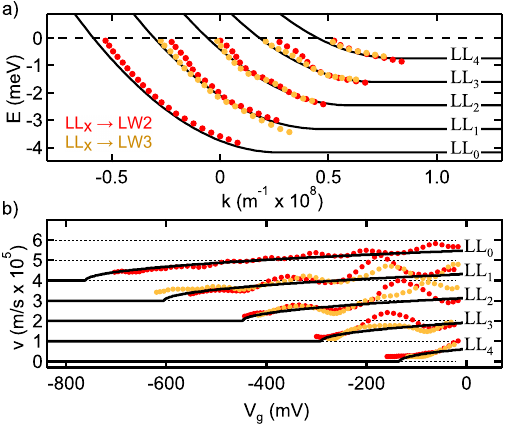}
	\caption{\textbf{Dispersions and velocities} (a) Dispersion of LLs extracted from the resonances in \ref{fig:3}(a), with solid curves corresponding to theory for hard wall edge confinement. Marker colors are the same as in \ref{fig:3}(a) and indicate tunneling from $\mathrm{LW_2}$ (red) and $\mathrm{LW_3}$ (orange). (b) Extracted velocities as a function of gate voltage for the resonances shown in \ref{fig:3}(a). Each LL is offset by 1 x $10^5$ m/s for clarity. Solid curves are the derivative $\partial E / \partial k_x $ of the theory curves shown in (a). }
	\label{fig:4}
\end{figure}
In the gate dependent measurement no direct energy scale is acquired due to the fixed bias voltage. Therefore, a different approach has to be considered. The energy of an electron at the Fermi level depends on the density which is varied by the gate voltage, such that $E=E(n(V_g))$. The velocity can then be expressed as,
 \begin{align}\label{eq:1}
 v = \frac{1}{\hbar} \frac{\partial E}{\partial k_x} &= \frac{1}{ed}\frac{\partial E}{\partial n }  \frac{\partial n }{\partial V_g} \frac{\partial V_g }{ \partial B_Y} ,
 \end{align}
The three derivatives are: $\partial E/ \partial n =D(E)^{-1}$, the inverse density of states (DOS), $\partial n / \partial V_g = \alpha$ the density lever arm, and $\partial V_g / \partial B_Y $  is the slope of the resonances as identified in Fig. \ref{fig:3}(a). Here, we have used the momentum kick $\Delta k_x=eB_Yd/\hbar$, and the relation $n(V_g)=\alpha (V_{g}-V_{th})$ where $V_{th}$ is the threshold voltage for the 2DEG depletion. 

In the presence of an out-of-plane magnetic field the constant DOS $D_{2D}$ is transformed into the comb of LLs. To perform the analysis we assume the constant DOS. This approximation becomes progressively worse as $B_Z$ is increased, but, as we will see, still yields good results at the present out-of-plane field of 0.5\,T. The resulting velocities are displayed in Fig. \ref{fig:4} (b) as a function of gate voltage. Two observations can be made. The velocity of the lower lying LLs is larger compared to higher LLs, and for all LL the velocity decreases as density is lowered and the LL is depleted. Both observations are consistent with a simple LL edge state picture as shown in Fig.\ref{fig:4} (a). The LLs closer to the edge bend up more strongly, obtaining a larger velocity. Further, we observe an oscillatory effect on the velocities, which can be attributed the simplified DOS approximation. 

Further we can convert the gate voltage into an energy by integrating \(\partial E / \partial V_g = -\alpha/D(E)\). With this, we can transform the extracted resonances of the form $V_g(B_Y)$ to the LL dispersion relation, $E(k_x)$, shown in Fig. \ref{fig:4} (a) for the first 5 spin degenerate LLs. Additionally, we show the calculated dispersions (solid curves) modeled from \cite{Patlatiuk2020} for edge states of a LL confined by a hard wall. The effect of the constant DOS used to extract the dispersions is much smaller here than for the velocities, because these depend on the derivative of the dispersions.  


In conclusion, we have employed two versions of momentum resolved tunneling to extract dispersions and velocities of edge states in the integer quantum Hall effect. First, using a small bias voltage we extracted the velocities over a large range of perpendicular magnetic fields. We observe a reduction of the velocities as the magnetic field is increased, in agreement with a simple hard wall model. Second, we performed spectroscopy while varying the density, thereby obtaining the evolution of the velocities at a fixed $B_Z$. In this way, we can directly access the edge state dispersions, and can even study the edge states of the same filling factors at many combinations of out-of-plane field and density.

In the future, this method can be applied to different material systems, such as HgTe, where intricate topological phases can occur \cite{Konig2007}. But even in the current device, these methods can help to shed more light on the complexities of the edge structure of fractional quantum Hall edge states \cite{MacDonald1990,Sabo2017}, since momentum resolved tunneling can give access to these complicated, sometimes even counter propagating edge modes. Similarly, one could potentially also discriminate between different 5/2 candidate states \cite{Manna2024}. As density is lowered, the ratio of kinetic to potential energy decreases, which can give rise to correlated phases \cite{Yoo2020, Pan2014} that may also be observed in their edge structure \cite{Lafont2019}. 
\\
\begin{acknowledgments}
We thank G. Barak for providing access to this sample. This research was funded by Swiss SNSF (grant no. 179024 and 215757) and the NCCR SPIN of the Swiss SNSF, and the EU H2020 European Microkelvin Platform EMP (Grant No. 824109). CPS acknowledges support by the Georg H. Endress Foundation.
\end{acknowledgments}

\bibliography{henoks_refs_merge3}

\end{document}


\title{Supplementary information\\ "Edge State Selective Measurement of Quantum Hall Dispersions"}
\maketitle
\tableofcontents
\clearpage
\section{Effect of finite size of the tunnel junctions} \label{sec:appendix_A}
\begin{figure*}[htb!]
    \includegraphics[width=0.8\textwidth]{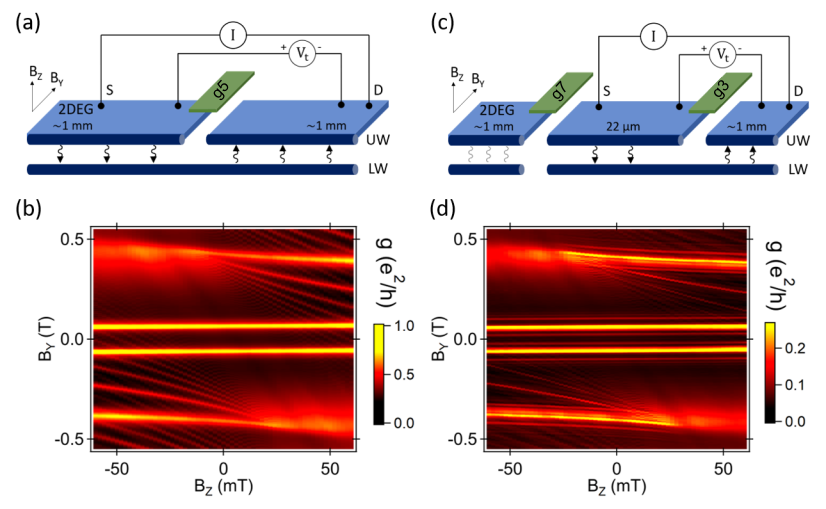}
    \caption{(a, b) Sample schematic and tunnel conductance measured as a function of in-plane field $B_Y$ and out-of-plane magnetic fields $B_Z$ with millimeter long source and drain tunnel junctions. (c, d) Analogous schematic and conductance map for the measurements with one short and one long tunnel junction.
    }
    \label{fig:a1}
\end{figure*}
The finite length of the tunnel junctions influences the appearance of the tunneling signal. To visualize the differences, we compare here the tunneling conductance measured using two long junctions versus the tunneling conductance measured with one long and one short junction.

The circuit diagram of the device tuned to the configuration with millimeter long source and drain tunnel junctions is shown in the Fig.\,\ref{fig:a1}(a). The tunneling conductance measured as a function of both in-plane field $B_Y$ and out-of-plane magnetic field $B_Z$ is shown in the panel (b), where two horizontal lines correspond to the $UW_1$-$LW_1$ tunneling resonances, and two fans with point symmetry about zero magnetic field are formed by the resonances involving tunneling from the Landau level edge states.

For the second measurement, the width of the source tunnel junction was reduced to 22\,$\mathrm{\mu m}$, see schematic in panel (c). The tunneling conductance measured in this configuration is shown in the panel (d). Multiple copies of the tunneling resonances with varying strengths appear in this conductance map. Though all the copies together are significantly broader than the single feature of the large junction, each copy itself is much more sharp and well-defined than the large junction resonance. In principle this makes it possible to track the exact position of these resonances much more precisely, though in some cases with multiple weaker copies nearby it can become difficult to assign to which state a particular replica belongs. This complicates the identification of some higher Landau level edge states. We note that both the short and long junctions happen to include some gates, which were grounded for these measurements, which may give rise to regions with slightly different densities. The short junction samples over a much smaller set of disorder sources and density variations along the edge, thus giving a discrete family of well-defined resonances. 
\clearpage
\section{Voltage correction}

\begin{figure*}[htb!]
    \includegraphics[width=1\textwidth]{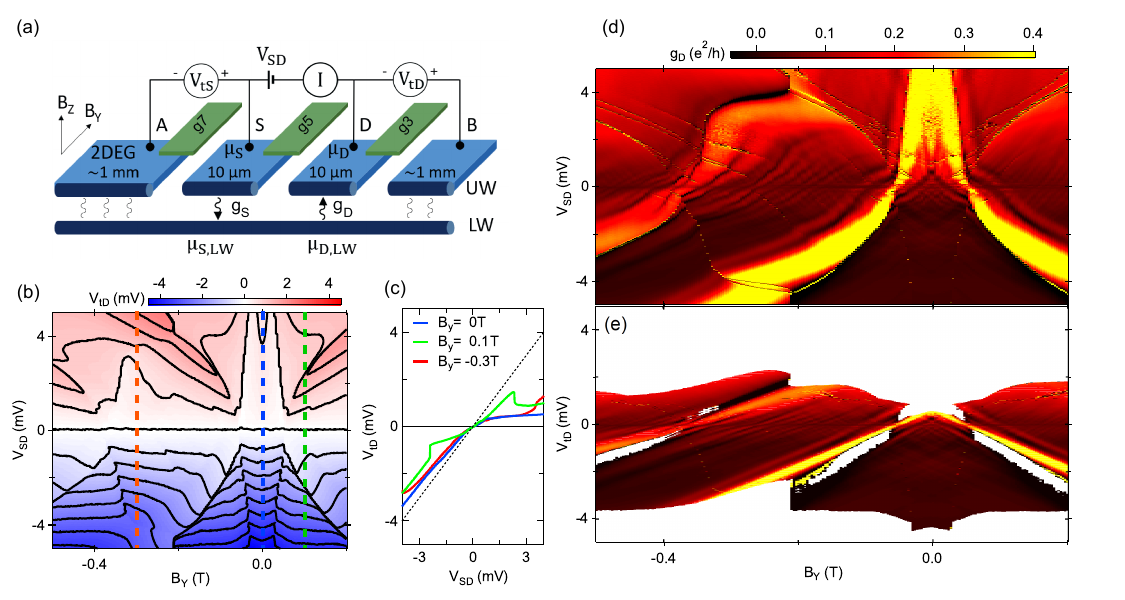}
    \caption{ (a) Schematic for the setup used for the tunneling measurement with simultaneous voltage measurement. (b) Measured drain junction voltage $V_{tD}$ as a function of applied $V_{SD}$ and spectroscopy field $B_Y$, highlighting the nonlinear behaviour. (c) Linecuts along the vertical dashed lines in (b), highlighting the assymetry between positive and negative $V_{SD}$. (d) Measured tunnel conductance on the drain junction as a function of $V{SD}$. (e) Same as (d) but now displayed as a function of the measured drain voltage $V_{tD}$.    }
    \label{fig:a2}
\end{figure*}

The source-drain voltage $V_{SD}$ gives access to the energy axis, but it drops over two tunnel junctions comprising several elements, see Fig.~\ref{fig:a2} (a). Thus, we need to separately measure the voltage drop over the tunnel junction of interest. For this, we use the additional voltage probes $A$ and $B$, as already described, providing the tunnel junction voltage $V_{tD}=-(\mu_{D,LW}-\mu_D)/e$ for the drain side and $V_{tS}=-(\mu_S - \mu_{S,LW})/e$ for the source side, where $e$ is the electron charge.

In Fig.~\ref{fig:a2} (b) we show the measured $V_{tD}$ as a function of $V_{SD}$ and $B_Y$. For positive $V_{SD}$, only a small part of the applied voltage $V_{SD}$ is dropping across the drain junction, indicated by the measured drain tunnel junction voltage $V_{tD}$ being well below the full applied voltage (dashed diagonal line), with the remaining voltage mainly dropping over the source tunnel junction.

We obtain the differential conductance $g_D=I_{AC}/V_{tD, AC}$, once as a function of $V_{SD}$, Fig.~\ref{fig:a2}  (d), and once as a function of $V_{tD}$, Fig.~\ref{fig:a2} (e).   While the two junctions are nominally identical, they are inverted with respect to the direction of current flow and voltage bias: for a negative $V_{SD}$, the source 2DEG is biased above the LW, while the drain 2DEG is biased below the LW. For negative $V_{SD}$ one can write:

\begin{equation} \label{eq1}
    \begin{split}
    \mu_{S} \geq \mu_{S,LW} \,\, \geq &\,\,  \mu_{D,LW} \geq \mu_{D} \equiv 0.  \\
    V_{tS}=-(\mu_S - \mu_{S,LW})/e \,\, , \,& -(\mu_{D,LW} - \mu_{D})/e=V_{tD} \\
    V_{SD} = V_{tS} + V_{tD}=-(&\mu_S - \mu_D)/e\,\,\,\,\,\,\mathrm{(ballistic)}
    \end{split}
\end{equation}

When $g_D$ is plotted as a function of tunnel junction voltage $V_{tD}$ instead, see Fig.\,\ref{fig:a2} (e), the previously highly curved stripes transform into nearly linear features with many parallel lines. For example, the wide yellow bands around $B_Y=0$ in panel (d) are straightened into the narrow yellow resonances following a line in panel (e). The resonances are related to the dispersions, and their slopes are proportional to the velocities of the states involved in the tunneling. The white regions in panel (e) represent the values of $V_{tD}$ not reached at all during the $V_{SD}$-scan. Since the applied $V_{SD}$ is divided between the two junctions, each tunnel junction voltage $V_{tD}$ and $V_{tS}$ reaches only a fraction of $V_{SD}=V_{tD}+V_{tS}$. Though the two tunnel junctions form a highly nonlinear and complex device, the described technique allows us to separately probe the voltages dropped across each junction, thus giving us access to the dispersion relations of the states involved in the tunneling.

\clearpage
\section{Complementary data from the source tunnel junction} 
\begin{figure*}[htb!]
    \includegraphics[width=0.9\textwidth]{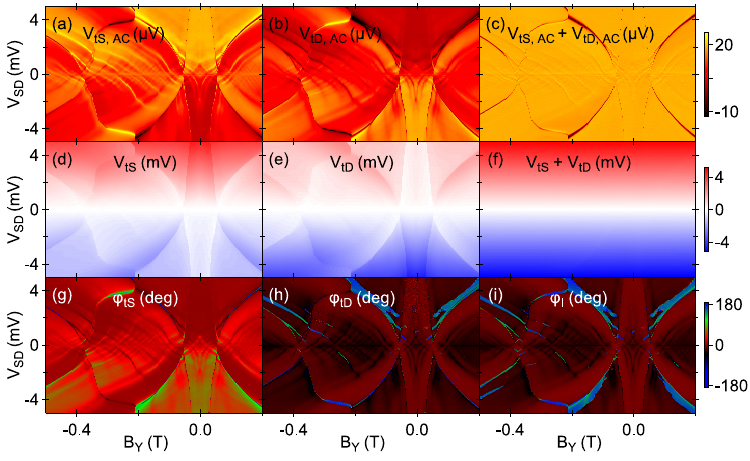}
    \caption{(a, b) The in-phase components of the AC voltage measured across the source and drain tunnel junctions and their sum shown in panel (c). (d, e, f) Analogous DC voltages. (g, h) The phase of the source and the drain AC voltages. (i) The phase of the AC component of the current.
    } \label{fig:a3}
\end{figure*}\label{sec:appendix_B}

For completeness, we also show all the AC and DC measurements in Fig.\,\ref{fig:a3}. The AC voltages measured across the source and the drain tunnel junctions are shown in Fig.\,\ref{fig:a3}(a) and Fig.\,\ref{fig:a3}(b), respectively. The sum of these two voltages is shown in panel (c), and for the majority of the points is very close to the AC voltage $V_{AC}=20~\mathrm{\mu V}$ applied to the device. Deviation appear only where the lock-in overloaded due to the abrupt current jumps, thus not delivering a valid measurement. Analogous DC measurements are shown in panels (d), (e), and (f). The sharp features present in the panels (d) and (e) compensate each other in the sum, see panel (f), which matches the full applied DC voltage $V_{SD}$ very well throughout the whole plot.  The phases of the source and drain AC voltages, $\phi_{tS}$ and $\phi_{tD}$, and the phase of the AC current, $\phi_{I}$, are shown in panels (g), (h), and (i), respectively. The voltage phase of the drain junction closely resembles the phase of the current, see panels (h) and (i). Due to the parasitic capacitance of the device and large resistance of the tunnel junctions, the phase of the source AC voltage, deviates substantially from the phase of the current. This deviation reduces upon lowering the measurement frequency (not shown here).

\clearpage
\section{Wire mode to wire mode tunneling} \label{sec:appendix_C}

\begin{figure*}[htb!]
    \includegraphics[width=0.85\textwidth]{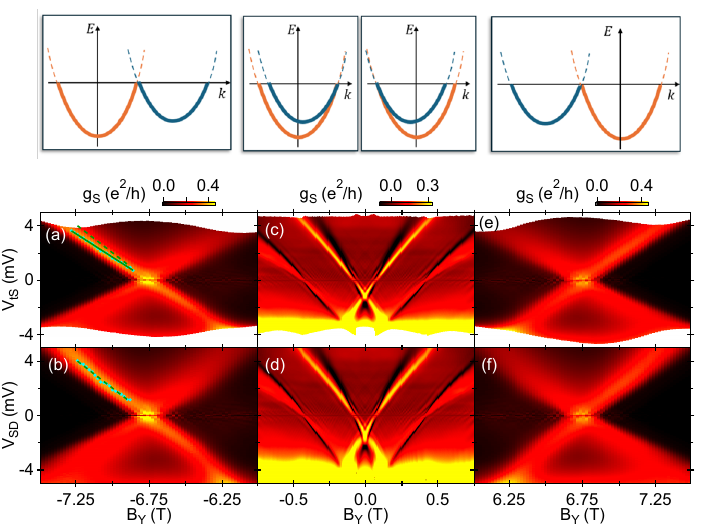}
    \caption{Differential tunneling conductance plotted as a function of $V_{tS}$, upper row, and $V_{SD}$, lower row, for the co-propagating, panels (c) and (d), and counter-propagating (a), (b), (e), (f) resonances. The measurements of the counter-propagating transition are repeated for positive and negative $B_Y$. The measurements were carried out at $B_Z = 0$ with the device tuned to the configuration with 6~$\mathrm{\mu m}$ long source and millimeter long drain tunnel junctions. The voltage drop across the source tunnel junction was measured using an additional ohmic contact.
    }
    \label{fig:a4}
\end{figure*}
The velocity of the charge excitations in a 1D wire is predicted to be elevated in the presence of strong electron-electron interactions. Such predictions of the Luttinger liquid model were observed in the quantum wire sample of Ref.~\cite{Auslaender2002} nominally identical to the one used in this study. In that study, the interaction strength was extracted from the ratio between the measured velocity and velocity calculated using a non-interacting Fermi gas model. The wire mode velocity was measured using momentum-resolved tunneling spectroscopy with the assumption that all the applied voltage is dropping across the narrow junction. This assumption leads to an overestimation of the excitation velocity, artificially indicating stronger interactions than actually present in the wires. Such a measurement is shown in panel (b) of Fig.~\ref{fig:a4}, for the case of right-moving electrons in the lower wire matching in momentum with left-moving electrons in the upper wire, i.e. one of the counter-propagating transitions. Similarly, we can plot this data against the measured source tunneling voltage $V_{tS}$, as shown in panel (a). The cross-shaped resonance formed by the two bright diagonal features map out the dispersion of the upper and lower wires close to the Fermi level. We note that the slightly distorted x-shape in (b) straightens and becomes more linear when plotted against the measured junction voltage in (a). The blue dots in both panels (a) and (b) indicate the position of the conductance maxima along the upper wire dispersion. These points were fit with a line, shown as solid green in panel (a) and dashed green in panel (b). In panel (a), we have also added the dashed line from panel (b) as a comparison to indicate the different slope resulting when the tunnel voltage is not separately measured. As seen, this line is clearly more steep, corresponding to a velocity which is a factor of $\sim1.2$ faster than with the measured tunnel junction voltage. Analogously, panels (e) and (f) show the conductance maps where the other pair of counter-propagating Fermi points are matched in momentum.

The panels (c) and (d) show the conductance maps with the matched co-propagating Fermi points. The narrow upper and wide lower white stripes in panel (c) indicate that $V_{tS}$ deviates substantially from $V_{SD}$ only for the negative bias voltage, leaving the upper part of the map (and the corresponding velocities) almost unchanged.

\clearpage
\section{Density spectroscopy as a function of out-of-plane magnetic field}
    \begin{figure*}[htb!] 
        \includegraphics[width=\textwidth]{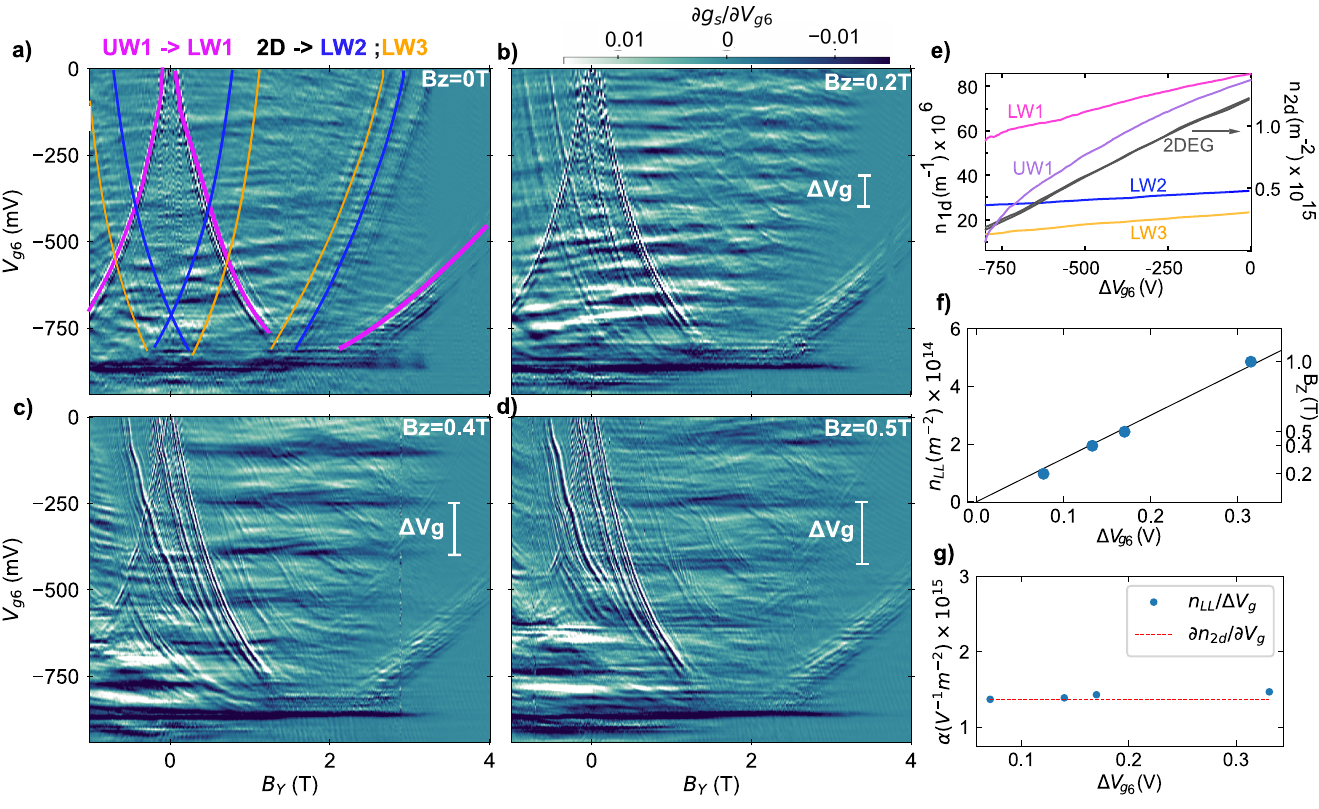}
        \caption{Spectroscopy maps for a range of $B_Z$. (a) $B_Z=0$, pink curves mark the transition between the first wire modes. Blue/yellow curves mark transitions between the 2DEG and $LW2$/$LW3$ respectively. (b-d) Measurements at increasing $B_Z$ as labeled. Horizontal features correspond to integer filling factors of the 2DEG. Their spacing is marked with $\Delta V_g$. (e) Densities extracted from the curves drawn in (a), showing the linear dependence of 2D density with agtevoltage. (f) $\Delta V_g(B_Z)$ displaying a linear relationship between $\Delta V_g$  and $n_{LL} = eB_Z/h$. (g) Lever arm extracted from (f) (blue dots) compared to lever arm from 2DEG properties (red dashed line) }
        \label{fig:a5}
    \end{figure*}
    
    Here we show the evolution of tunnel maps $g(V_{g6})$ for a range of increasing values $B_Z$. Fig. \ref{fig:a5}(a) was taken at $B_Z=0$ and displays tunneling between the first wire modes $UW1-LW_1$ shown in purple and tunneling between the 2DEG and higher $LW$ modes. The tunneling resonance between the 2DEG and $LW_2$/$LW_3$ is shown in blue/yellow. Following Auslaender et al.\cite{Auslaender2005} the Fermi wave-vectors of each of those modes can be extracted as a function of gate-voltage. At a given gate voltage, there are 2 magnetic fields where the momenta of the participating modes are matched, corresponding to the co-propagating and counter-propagating transitions $B_{Y_{ij}}^-$ and $B_{Y_{ij}}^+$, where the indices $i,j$ indicate the corresponding mode in the upper ($LL$ or $UW$-mode) and lower ($LW$-mode) system respectively. At these specific fields, the following equations are fulfilled:

    \begin{align}
        B_{Y_{ij}}^+=\hbar(k_{F_i}+k_{F_j})/ed \\
        B_{Y_{ij}}^-=\hbar(k_{F_i}-k_{F_j})/ed
     \end{align}

    Here $k_{F_i}$ and $k_{F_j}$ are the Fermi-momenta of the two involved modes. From these equations we arrive  at the wave vectors
    \begin{align}
       k_{F_i}=ed(B_{Y_{ij}}^++B_{Y_{ij}}^-)/2\hbar \\
       k_{F_j}=ed(B_{Y_{ij}}^+-B_{Y_{ij}}^-)/2\hbar.
     \end{align}
   Using the extracted Fermi wave-vectors the densities can be calculated as $n_{1d}=2k/\pi$  and $n_{2D}=k^2/2\pi$. The extracted densities are shown in Fig. \ref{fig:a5}(e). For the 2DEG, the density is extracted once for tunneling to each $LW_2$ and $LW_3$. Both extracted 2DEG densities agree with each other, and the density curves shown in grey in Fig. \ref{fig:a5} (e) lie on top of each other. In addition, we also show the density for the 1d wire modes for $UW_1$, $LW_1$, $Lw_2$, and $Lw_3$. It is noteworthy, that the first UW mode depletes faster than the first LW mode, due to screening.  
   
    The tunneling resonances with the 2DEG evolve into the $LL$ resonances. Fig. \ref{fig:a5} b), c), d) show the evolution for increasing $B_z$.  While the $UW_1-LW_1$ transition remains almost unchanged (except for the occurrence of hybridization gaps), transitions from the 2DEG are not easily visible any longer. Instead, they are replaced by resonances of tunneling between $LL$ and $LW_{2/3}$. 
    
    Other than the momentum-matched resonances which are both $B_Y$ and $V_g$ dependent, there is a $B_Y$  independent contribution, most likely stemming from the bulk $LL$ which is momentum independent, proportional to the DOS. In the derivative shown, this emerges as the horizontal dark lines. The gate voltage separation between two successive lines $\Delta V_g$ corresponds to the depletion of a spin degenerate $LL$. This is due to the Zeeman energy being much smaller than the cyclotron energy ($\hbar \omega_c/E_z \sim 70$ in GaAs). The resulting wave vector difference between spin split $LL$ is therefore quite small, and cannot be resolved. 
    
    The density associated with a $LL$ is $n_{LL}=h/2eB_Z$, where the factor of 2 accounts for spin. When $B_Z$ is increased the $LL$ degeneracy, and therefore, $n_LL$ increases. The relationship between $\Delta V_g$ and $B_Z$ is shown in Fig. \ref{fig:a5} (f). The linear relation shows that the density lever arm, i.e. the electron density depleted per unit gate voltage, is a constant, independent of $B_Z$ for the range studied. The density lever arm can also be obtained by taking the derivative of $n_{2D}(V_g6)$. Both are shown in Fig. \ref{fig:a5} (g) confirming the assumption of a magnetic field independent density lever arm. The slight slope of the data (blue dots) seen in Fig.\ref{fig:a5} (g) could indicate a slight change in density lever arm at larger magnet fields. In any case, this would be a small effect which does not affect our conclusions. More detailed measurements would be needed to come to a final conclusion on this.  

\clearpage
\section{Hybridization of Landau-level with upper-wire modes}\label{sec:appendix_D}
    \begin{figure*}[htb!]
        \includegraphics[width=0.7\textwidth]{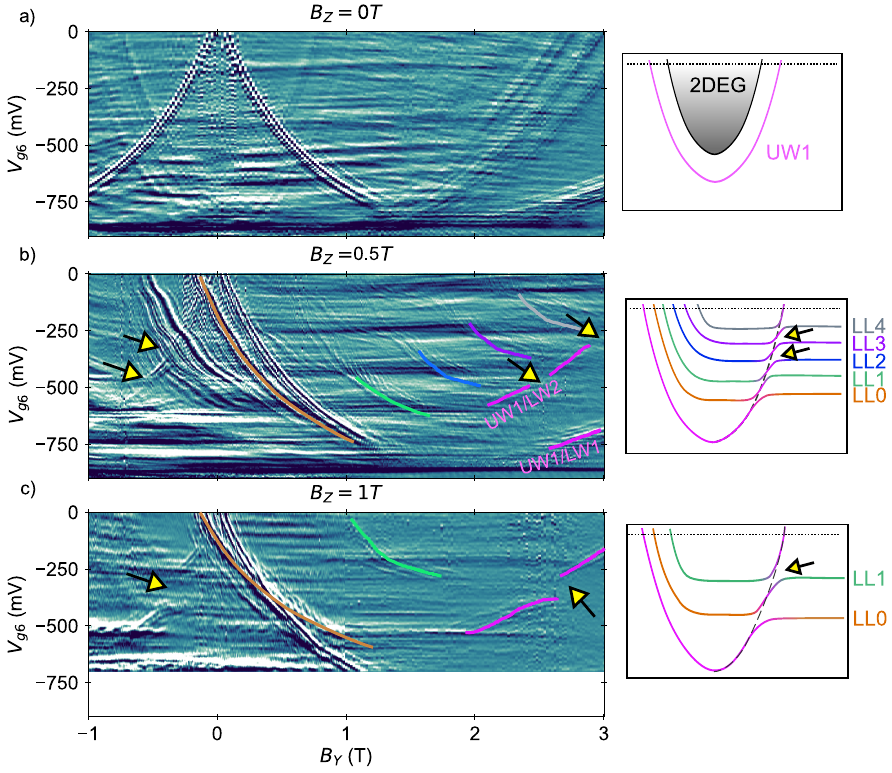}
        \caption{Avoided crossings between wire modes and Landau levels. The first column shows the measured data, while the second column is the corresponding sketch of the band structure at the respective out-of-plane field $B_Z$. (a) At $\mathbf{B_Z=0}$\,T there are no $LL$s present, so the wire-wire transition (around $B_Y=0$) remains continuous. The resonances are symmetric with respect to $B_Y$. (b) At $\mathbf{B_Z=0.5}$\,T $LL$ have formed.  Gaps in the $UW1-LW1$ transition from hybridization with $LL$ become visible for negative $B_Y$, indicate by yellow arrows, while the same resonance at postive $B_Y$ remains continuous. The avoided crossings can be observed above $B_Y=2$\,T, for the $UW1-LW2$ transition, also indicated by yellow arrows.  (c) Same as (b), but at $\mathbf{B_Z=1}$\,T. The size of the gap increases.   }
        \label{fig:a6}
    \end{figure*}  
    By means of momentum-resolved tunneling spectroscopy at fixed $B_Z$ field as a function of gate voltage and spectroscopy field, we can directly measure the structure of the hybridized system of $UW$ states and $LL$. This hybridization and the associated avoided crossings were previously predicted and observed indirectly \cite{Patlatiuk2020} in a measurement of $B_Z$ vs $B_Y$, where they appeared as discontinuities in the tunneling resonances. By keeping $B_Z$ fixed and changing the density instead, we are able to observe the avoided crossings of all $LL$ with the first $UW$ mode simultaneously.  This can be seen in Fig. \ref{fig:a6}, where we show density dependent spectroscopy measurements for different $B_Z$.  Figure \ref{fig:a6} (a) shows the situation at $B_Z=0$\,T where no $LL$ are formed. Therefore, no Hybridization occurs. Figure \ref{fig:a6} (b) and (c) are $B_Z=0.5$\,T and $B_Z=1$\,T respectively and display avoided crossings, most notable by the gaps along one side of the wire-wire transition (yellow arrows). 
\clearpage
\section{Numerical Models}


\subsection{Landau level velocity as a function $B_Z$ (Fig. 3)}
As in \cite{Patlatiuk2018} the Schrödinger equation is solved for a parabolic confinement due to the applied magnetic field, centered on the guiding center position, and cut off by  the presence of the hard wall close to the edge. By solving this problem numerically the dispersions for each magnetic field are obtained, from which the velocities can then be calculated. For this the slope $\partial E / \partial k_x $ is calculated at the Fermi-level, as obtainend from the 2DEG density. 

\subsection{Landau level dispersions and velocities as a function of $V_g$ (Fig. 4)}
Following \cite{Patlatiuk2018} the dispersion of the n\textsuperscript{th} LL, $E_n(k_x)$ can be written as 
\begin{equation}\label{eq:disp}
    E_n(k_x) =  E_n^{Bulk} +\frac{\hbar^2}{2m^*} \Theta(\sigma_n - Y) \left(\frac{\sigma_n}{l_B^2}-k_x\right)^2 . 
\end{equation}
Here, $E_n^{Bulk} = \hbar \omega_c (n+1/2)$  is the energy of the n\textsuperscript{th} LL in the Bulk, $\sigma_n\approx2l_B\sqrt{2n+1}$ is the width of the n\textsuperscript{th} LL, $l_B=\sqrt{\hbar/eB_Z}$ is the magnetic length, $\omega_c= eB_Z/m^*$, $Y= k_x l_B^2$ is the guiding center position, with the hard wall placed at $Y=0$, and $\Theta$ is the Heaviside function.  

To obtain the shown traces in 4b) for the velocities as a function of gate voltage, we obtain the density $n$ at the respective gate voltage. From this we calculate the Fermi energy and find the crossing points of the dispersion of the respective Landau level with the Fermi level where $E_n(k_x)=E_F$. We then calculate the velocity $v_n=\partial E_n(k_x) /\partial k_x|_{k_x=k_F}$ and repeat these steps at each gate voltage.
\bibliography{henoks_refs_merge3}